\documentclass[letterpaper,prl,twocolumn,superscriptaddress]{revtex4}
\usepackage{graphicx,bm}

\begin{document}
\title{Edge magnetism of Heisenberg model on honeycomb lattice}
\author{Wen-Min Huang}
\email{wmhuang0803@gmail.com}
\affiliation{Department of Physics, National Tsing Hua University,
30013 Hsinchu, Taiwan}
\affiliation{Physics Division, National Center for Theoretical Sciences, Hsinchu 300, Taiwan}
\author{Yen-Chen Lee}
\affiliation{Department of Physics, National Tsing Hua University,
30013 Hsinchu, Taiwan}
\author{Toshiya Hikihara}
\affiliation{Faculty of Engineering, Gunma University, Kiryu, Gunma 376-8515, Japan}
\author{Hsiu-Hau Lin}
\email{hsiuhau@phys.nthu.edu.tw}
\affiliation{Department of Physics, National Tsing Hua University,
30013 Hsinchu, Taiwan}
\affiliation{Physics Division, National Center for Theoretical Sciences, Hsinchu 300, Taiwan}

\date{June 13, 2011}
 
\begin{abstract}
In our previous study, a single-branch of ferromagnetic magnon with linear dispersion is shown to exist near the (uncompensated) zigzag edge for Heisenberg model on honeycomb lattice. Here we develop a field-theory description for the edge magnon and find its dynamics is captured by the one-dimensional relativistic Klein-Gordon equation. It is intriguing that the boundary field theory for the edge magnon is tied up with its bulk counterpart, described by the two-dimensional Klein-Gordon equation. Furthermore, we also reveal how the parity symmetry relates evanescent modes on opposite edges in a honeycomb nanoribbon. By employing alternative methods, including Schwinger bosons and density-matrix renormalization group, we also demonstrate that the relativistic edge magnon is robust even when the Neel order in the bulk is destroyed by quantum fluctuations. The edge magnon is a direct consequence of uncompensated edge and may be verified in realistic materials by experimental probes.
\end{abstract}

\maketitle

\textit{Introduction.}
Exchange interactions between local magnetic moments, often described by Heisenberg model\cite{Heisenberg1926} and its derivatives, lead to rich and sometimes exotic phases in quantum magnetism\cite{Auerbach2006}. For instance, the excitation gap in the integer-spin chain proposed by Haldane\cite{Haldane1983} stimulates theoretical investigations and is later verified in experiments\cite{Buyers1986,Renard1987}. Antiferromagnetism in two-dimensional square lattice has been studied extensively because of its adjacency to unconventional superconductivity in cuprates\cite{Manousakis1991} and iron-based materials\cite{Paglione2010}. Recent breakthrough shows that $S=1/2$ Heisenberg model on Kagome lattice exhibits exotic spin-liquid ground state\cite{Yan2011} due to strong quantum frustrations. Moreover, it has been demonstrated that superexchange interactions between ultracold atoms can be realized in optical lattices\cite{Trotzky2011}. It may provide a different route to understand various ground states of the Heisenberg model on different lattice structures.

It is known that boundary effects give rise to fractionalized excitations in integer-spin chains \cite{Hagiwara1990,Glarum1991} but are less studied for spin systems in higher dimensions. The importance to understand the boundary effects in Heisenberg model is echoed by potential edge magnetism in graphene nanoribbon\cite{Fujita1996,Wakabayashi1998,Wakabayashi1999,Hikihara2003,Fernandez-Rossier2007,Jiang2008,Yazyev2010}. Monte Carlo simulations\cite{Feldner2011} demonstrate mean-field like ferromagnetic moment near the zigzag edge of graphene nanoribbon. Unlike the usual ferromagnetic magnons with quadratic dispersion, these sharp spin excitations near zigzag edge shows linear dispersion. Recent spin-wave calculations\cite{You2008} for the Heisenberg model on honeycomb lattice show that the dispersion of the ferromagnetic edge magnon is indeed linear.

\begin{figure}
\begin{center}
\includegraphics[width=7.5cm]{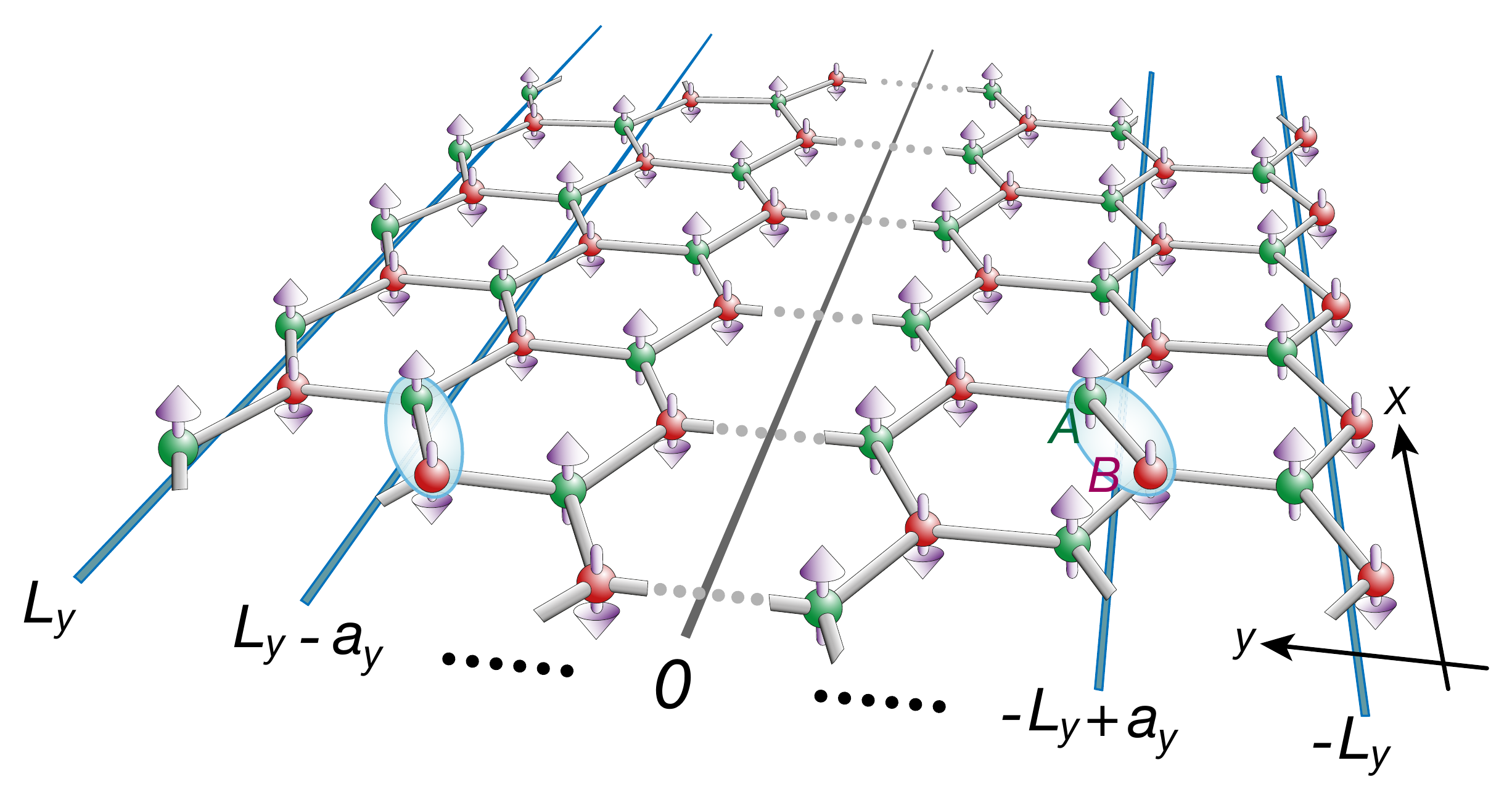}
\caption{Neel state in a honeycomb nanoribbon with zigzag edges. Spin orientations (purple arrows) on sublattice $A$ (green dots) and sublattice $B$ (red dots) are opposite to each other. The choice of unit cell is highlighted by the shaded blue circle.
}
\label{Ribbon}
\end{center}
\end{figure}

The emergence of edge magnetism can be understood by Lieb's ferrimagnetism. However, it remains puzzling how the ferromagnetic edge magnon acquires linear dispersion, which is clearly related to the antiferromagnetic correlations in the bulk. In fact, the absence of quadratic dispersion hints that the boundary field theory for the edge magnon must tie up with the bulk, just like the chiral states in quantum Hall effects. In this Letter, we study edge magnetism of the Heisenberg model on honeycomb nanoribbon with zigzag edges as shown in Fig. 1. Within spin-wave approximation, magnons in the bulk follow the two-dimensional Klein-Gordon (K-G) equation, characterized by the spin-wave velocity $c_b$ and the effective mass $m_b$. The boundary condition gives rise to evanescent modes and imposes constraint on the propagation along the edge and the (imaginary) momentum in the transverse direction. It is quite remarkable that the edge magnon is described by the one-dimensional K-G equation with parameters $c_e$ and $m_e$ related to their bulk values. Our derivations provide natural explanation for the linear dispersion and reveal the connection between the boundary field theory for edge magnons and its bulk counterpart. Though the above conclusions are drawn from spin-wave theory, we further extend the calculations by Schwinger boson approach, where the bulk no longer has long-ranged Neel order, and find the edge magnon exhibit qualitatively the same behavior. Meanwhile, we also perform density-matrix renormalization group (DMRG) calculations to demonstrate that the spin-singlet ground state in nanoribbon is actually very close to the Neel state.

\textit{Harper equations.}
To explore the boundary effects for Heisenberg model on honeycomb nanoribbon, we first write down the Hamiltonian for exchange interactions,
\begin{eqnarray}
\nonumber H&&\hspace{-0.3cm}=\sum_{\langle\bm{r},\bm{r}'\rangle} \sum_a
J_a\left(\bm{r},\bm{r}'\right)\hspace{0.1cm} S_a\left(\bm{r}\right) S_a\left(\bm{r}'\right),
\end{eqnarray}
where $\langle\bm{r},\bm{r}'\rangle$ is denoted the nearest-neighbor pairs and $J_x=J_y=J$, $J_z=\gamma J$ are exchange couplings with anisotropy $\gamma\geq1$. The spin operators can be represented by Holstein-Primakov (HP) bosons. Assuming the ground state has Neel order as shown in Fig.~\ref{Ribbon}, the interactions between these bosons can be ignored and the effective Hamiltonian within spin-wave approximation is
\begin{eqnarray}
&&\hspace{-0.8cm}\nonumber H_{SW}=JS\sum_{\langle\bm{r},\bm{r}'\rangle}\bigg[\gamma\left(b^{\dag}_A({\bm r})b_A({\bf r})+b^{\dag}_B({\bm r}')b_B({\bm r'})\right)\\&&\hspace{2cm}+b^{\dag}_A({\bm r})b^{\dag}_B({\bm r}')+b_B({\bm r}')b_A({\bm r})\bigg],
\end{eqnarray}
where $S$ is the magnitude of the spin and $b_{A/B}$ are annihilation operators for HP bosons on sublattices $A/B$. Since the spin-wave Hamiltonian is bilinear in HP bosons, it is more convenient to write down its equivalent equations of motion in first-quantization language. Following the same steps developed in Ref.~\cite{You2008}, the dynamics is described by the coupled Harper equations,
\begin{eqnarray}
&&\lambda\varphi_A(\bm{r})+Q\sum_{\bm{\delta}_i}\varphi_B^*(\bm{r}+\bm{\delta}_{i})=i\partial_t \varphi_A(\bm{r}),\\
&&Q\sum_{\bm{\delta}_i}\varphi_A(\bm{r}+\bm{\delta}_{i})+\lambda\varphi_B^*(\bm{r})=-i\partial_t\varphi_B^*(\bm{r}),
\end{eqnarray}
where $\bm{r}$ denotes the lattice sites for honeycomb lattice and $\bm{\delta}_{i}$ are the vectors pointing to the nearest neighbors. The wave functions on different sublattices are $\varphi_A, \varphi^*_B$, where the conjugation arises from opposite spin orientation. The parameters are $Q=JS$ and $\lambda=z\gamma Q$, where $z$ is the number of nearest neighbors. It is important to emphasize that the magnon carries quantum number $\Delta S_z=\mp1$ and sets the normalization condition,
\begin{eqnarray}
\Delta S_z=-\sum_{\bm{r}}\left(\left|\varphi_A(\bm{r})\right|^2-\left|\varphi_B(\bm{r})\right|^2\right)=\mp1
\end{eqnarray}
The above Harper equations can be solved exactly, delivering a single-branch ferromagnetic magnon near the zigzag edge with linear dispersion. In the following, we would like to develop general field-theory descriptions to explicitly reveal the connection between magnons in the bulk and those at the edge.

\textit{Field theory for the bulk.}
In the field-theory limit, we introduce the smooth-varying fields, $\phi_{\Lambda}(\bm{r},t) = \frac{1}{V} \sum_{|\bm{k}|<\Lambda_c}\varphi_{\Lambda}(\bm{k},t)\: e^{i \bm{k}\cdot\bm{r}}$, where the momentum summation is restricted to the vicinity of $\bm{k}=0$ with a cutoff $\Lambda_c$. For these smooth-varying fields, spatial variable $\bm{r} = (x,y)$ can be treated as continuous and no longer restricted to the lattice sites. As a consequence, spatial derivatives are well-defined.  Making use of the displacement operator, $e^{\bm{a} \cdot \nabla} \phi(\bm{r}) = \phi(\bm{r}+\bm{a})$, the Harper equations can be represented in the matrix form,
\begin{eqnarray}\label{SE}
\left[\begin{array}{cc}\lambda & Qh(\partial_x,\partial_y) \\Qh(\partial_x,-\partial_y) & \lambda\end{array}\right]\left[\begin{array}{c}\phi_A \\\phi_B^*\end{array}\right]=i\partial_t\left[\begin{array}{c}\phi_A \\-\phi_B^*\end{array}\right],
\end{eqnarray}
where $h(\partial_x,\partial_y)=2e^{-\sqrt{3}a\partial_y/6}\cosh(a\partial_x/2)+e^{\sqrt{3}a\partial_y/3}$ for honeycomb lattice. Keeping the lowest order in gradient expansions and eliminating the field $\phi^*_B$, the dynamical equation solely for the field $\phi_A$ can be derived. It is not surprising that the effective field theory turns out to be the well-known Klein-Gordon equation in two dimensions,
\begin{eqnarray}
\left[\frac{1}{c_b^2} \frac{\partial^2}{\partial t^2} 
- \left( \frac{\partial^2}{\partial x^2} + \frac{\partial^2}{\partial y^2}\right) 
+ m_b^2 c_b^2 \right] \phi_A(x,y,t) = 0.
\end{eqnarray}
The spin-wave velocity in the bulk is $c_b= \sqrt{3/2}\:Qa$ and the effective mass is $m_b=2\sqrt{\gamma^2-1}/(Qa^2)$. One can also eliminate the field $\phi_A$ and show that
$\phi_B^*$ also satisfies the same K-G equation. These results are not surprising because antiferromagnet in low-energy limit is relativistic. Without the annoying spin kinematics, spin operators can be viewed as canonical bosons and K-G equation is a natural description for relativistic bosons.

It is important to keep in mind that $\phi_A$ and $\phi^*_B$ are indeed antiparticles to each other. As required by relativity, they always appear in pairs and explain the double degeneracy for magnons in an antiferromagnet. Furthermore, when the anisotropy disappears, $\gamma=1$, the excitation gap for the magnon $m_b c_b^2$ also disappears as expected from Goldstone's theorem.

\textit{Field theory for the edge.}
One can also introduce the smooth-varying fields on the edge and applies the same techniques to derive the boundary field theory. Since our goal is to demonstrate the connection between the field theories in the bulk and on the edge, it is wise to write down the field-theory presentation for the boundary conditions. For the honeycomb nanoribbon considered here, at the upper edge ($y=L_y$) where the outmost sites belong to sublattice $A$, the boundary condition gives the constraint $\gamma\varphi_A(x,L_y) + \varphi^*_B(x,L_y+a_y) = 0$. On the other hand, for the lower edge ($y=-L_y$), the outmost sites belong to sublattice $B$ and the boundary condition leads to $\gamma\varphi_B(x,-L_y) + \varphi^*_A(x,-L_y-a_y) = 0$. As long as the transverse width $L_y$ is finite, edge magnons on opposite edges entangle together and complicate the problem. For simplicity, let us temporarily assume that the transverse width $L_y$ is sufficiently large so that the coherent overlap between opposite edges can be ignored. 

Eliminating the field $\varphi^*_B$ with the help of Eq.~(\ref{SE}), the boundary conditions on the upper edge is simplified to the constraint on the field $\varphi_A$ solely,
\begin{eqnarray}\label{BC1}
\frac{\partial \phi_A}{\partial y} - i \left(\frac{1}{\sqrt{2}\gamma}\right) \frac{1}{c_b}\frac{\partial \phi_A}{\partial t} =0.
\end{eqnarray}
Note that the above relation impose constraint on how the (imaginary) momentum in the transverse direction renormalizes the propagation of magnons on the upper edge. It can be shown that edge magnons on the upper boundary carry quantum number $\Delta S_z=-1$ with evanescent wave function $\phi_A(x,y,t) = \phi_{Ae}(x,t) e^{\alpha_y y}$, where $\alpha_y>0$ is the imaginary momentum along the transverse direction. Substituting the boundary constraint into the bulk K-G equation, the dimensionality is effectively reduced to one. The resultant equation for edge magnon is the one-dimensional K-G equation,
\begin{eqnarray}
\left[\frac{1}{c_e^2} \frac{\partial^2}{\partial t^2} 
- \frac{\partial^2}{\partial x^2} 
+ m_e^2 c_e^2 \right] \phi_{Ae}(x,t) = 0.
\label{1d-KG}
\end{eqnarray}
The spin-wave velocity and the effective mass for the edge magnon are related to its bulk values,
\begin{eqnarray}
\frac{c_e}{c_b} = \frac{m_b}{m_e} = \frac{1}{\sqrt{1+\frac{1}{2\gamma^2}}} <1.
\end{eqnarray}
The above results are plotted in Fig. 2. Since the excitation gap $m_e c_e^2$ and the spin-wave velocity $c_e$ are smaller, the dispersion for the edge magnon lies below the continuum and remains sharp even when the interactions between magnons are included perturbatively.

\begin{figure}
\begin{center}
\includegraphics[width=8.2cm]{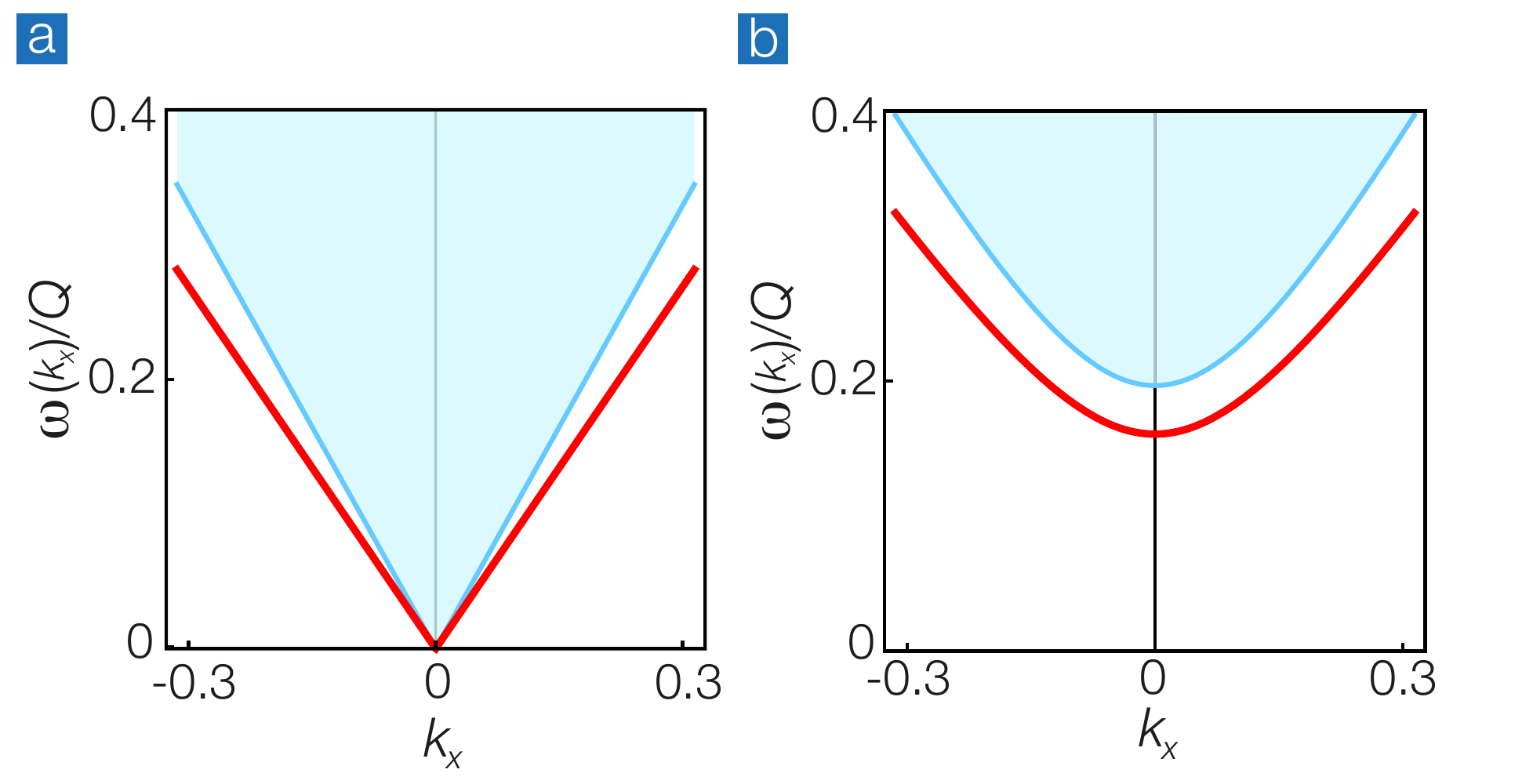}
\caption{Dispersions for edge and bulk magnons, (a) in isotropic limit $\gamma=1$ and (b) with slight anisotropy $\gamma=1.01$. The dispersion of edge magnon shows linear dependence as indicated by red lines. The shaded light blue regime represents the continuum of magnons in the bulk.}
\label{dispersion}
\end{center}
\end{figure}

On the lower boundary, the edge magnons carry quantum number $\Delta S_z=1$ with wave function $\phi_B^*(x,y,t) = \phi_{Be}^*(x,t) e^{-\alpha_y y}$. Following similar calculations, one can show that the edge magnon also satisfies the one-dimensional K-G equation with identical parameters. The similarity between the edge magnons on opposite edges calls for a  $Z_2$ symmetry argument. It turns out the discrete symmetry relating these evanescent modes is the parity symmetry $P_y$ in the transverse direction.

Because the Neel state has staggered spin configuration, the operation of parity symmetry needs extra caution. When reversing the $y-$axis, it is clear that the lattice coordinates transform as $(y,A) \to (-y, B)$ and $(y,B) \to (-y,A)$. However, due to the staggered spin configuration in the Neel state, the parity transformation also reverses the spin orientations and causes ``charge conjugation" effectively. That is to say, the parity transformation turns particle-like excitations into the hole-like and vice versa. Therefore, the solution under $P_y$ transformation takes the form, 
\begin{eqnarray}
 \left[\begin{array}{c}
\phi_A(x,y,t)
\\
\phi^*_B(x,y,t)
\end{array}\right]
\rightarrow
\left[\begin{array}{c}
\phi^*_B(x,-y,t)
\\
\phi_A(x,-y,t)
\end{array}\right].
\end{eqnarray}
The above symmetry is exactly what happens in the boundary field theory for edge magnons.

The boundary field theory supplemented with the symmetry argument fully answers our puzzle. The ferromagnetic magnons satisfies the one-dimensional K-G equation originated from its bulk counterpart with explicit relations. The edge magnons running on upper and lower boundaries carry opposite quantum numbers and are antiparticles to each other related by the $P_y$ parity symmetry. In fact, the whole field theory (including the bulk and the two edges) is fully relativistic and excitations always appear in pairs as required. The confusion mainly arises from the asymmetry of the spatial wave functions for the edge magnons because their antiparticles locate on the opposite edges. In short, the single-branch {\em ferromagnetic} edge magnon on one zigzag boundary is indeed an {\em antiferromagnetic} one with its antiparticle running on the distant opposite boundary. The linear dispersion of the edge magnon (with specific relation to the bulk dispersion) now looks more than natural.

\textit{General cases.}
One may wonder how robust the above conclusions are if the Neel order in the bulk is absent.
To explore this possibility, we represent the spin operators by Schwinger bosons which preserve the $SU(2)$ symmetry in the isotropy limit $\gamma=1$ explicitly. Since the calculations are pretty much the same as the HP bosons, we skip the technical details here. We arrive at the same conclusions except the parameters $Q$ and $\lambda$ in the Harper equations are modified due to quantum fluctuations. The robustness of the relativistic boundary field theory should not be a big surprise because the field-theory description for Schwinger bosons is still relativistic.

\begin{figure}
\begin{center}
\includegraphics[width=8.2cm]{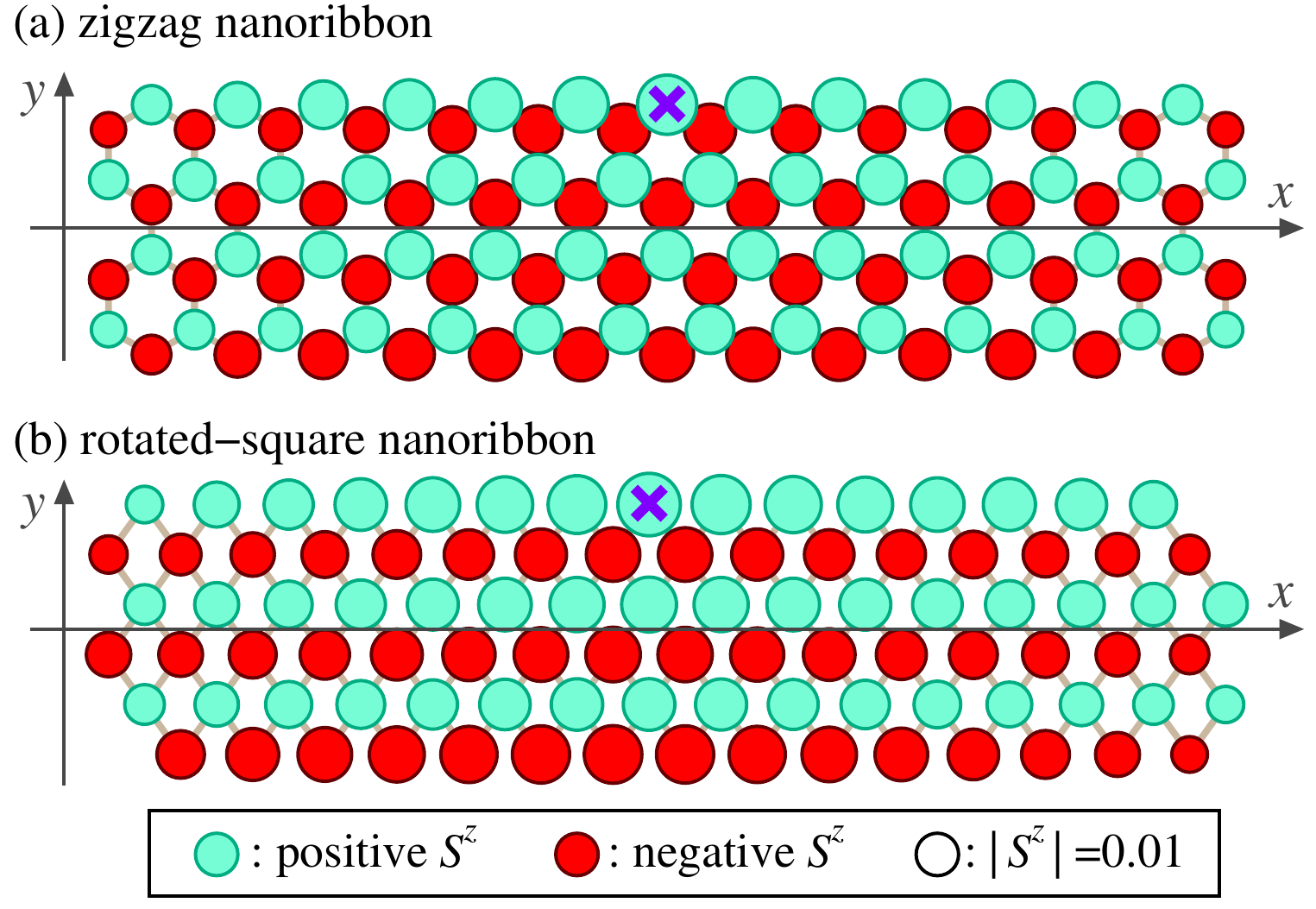}
\caption{Magnetization profiles $\langle S_z({\bf r}) \rangle$ for (a) in zigzag nanoribbon with $L_y = 2$ and (b) in rotated-square nanoribbons with $L_y = 3/2$. Light (green) and dark (red) circles represent positive and negative values of the spin polarization respectively, while the areas are proportional to the absolute values. Crosses represent the edge site ${\bf r}_0$ for which the local Zeeman field $-h S_z({\bf r}_0)$ with $h = 0.01J$ is applied.}
\end{center}
\end{figure}

We also perform DMRG calculations to demonstrate that the ground state is very close to the Neel state. It is known that the ground state for Heisenberg model on finite bipartite lattice is a spin singlet. Figure 3 shows the magnetization profiles $\langle S_z({\bf r}) \rangle$ in the lowest-energy state of $\sum_{\bf r} S_z({\bf r}) = 0$ for the Heisenberg model with local Zeeman field $-h S_z({\bf r}_0)$ applied to a center site ${\bf r}_0$ of one edge. The result suggests that even the small local field $h=0.01J$ turns the ground state to almost perfect Neel state. It is also interesting to notice that the edge magnon is not a privilege of honeycomb nanoribbon with zigzag edges. For rotated-square nanoribbon as shown in Fig. 3, we repeat the same calculations and find the presence of edge magnon as described in Eq.~\ref{1d-KG}, with different parameters,
\begin{eqnarray}
\frac{c_e}{c_b} = \frac{m_b}{m_e} = \frac{1}{\sqrt{1+\frac{1}{\gamma^2}}} <1,
\end{eqnarray}
where $c_b = 4Qa$ and $m_b = \sqrt{\gamma^2-1}/(4Qa^2)$. Therefore, the emergence of the edge magnon is related to the uncompensated lattice structure.

Although the Neel order may not play an essential role to the robustness of the edge magnon, one shall be cautious to draw similar conclusions for graphene nanoribbon with the same geometry. Since edge magnetism in graphene nanoribbon is itinerant in nature, it is not yet clear whether the edge magnon can still be described by the one-dimensional K-G equation derived here. However, recent Monte Carlo simulations\cite{Feldner2011} demonstrate the presence of sharp spin-wave excitation with sharp spectral weight, in qualitative agreement with our boundary field-theory description. It would be of vital importance to explore and reveal the true nature of these edge excitations in the future. 

We acknowledge supports from the National Science Council in Taiwan through grant NSC-97-2112-M-007-022-MY3 and T.H. was supported in part by the MEXT and JSPS, Japan through Grand No. 21740277. Financial supports and friendly environment provided by the National Center for Theoretical Sciences in Taiwan are also greatly appreciated.

\end{document}